\documentclass[aps,prc,superscriptaddress,twoside,twocolumn,nofootinbib,showpacs]{revtex4}
\usepackage{amsmath,amssymb}
\usepackage{graphicx}
\newcommand*{\fs}[1]{{#1\!\!\!/}}
\newcommand*{\hc}{\text{H.\,c.}}
\begin{document}

\title{\boldmath Analyzing $\eta'$ photoproduction data on the proton at energies of 1.5--2.3~GeV}
\author{K. Nakayama}
\affiliation{Department of Physics and Astronomy, University of Georgia, Athens, GA 30602, USA}
\affiliation{Institut f{\"u}r Kernphysik (Theorie),
Forschungszentrum J{\"u}lich,
D-52425 J{\"u}lich, Germany}
\author{H. Haberzettl}
\affiliation{Center for Nuclear Studies,
Department of Physics,
The George Washington University,
Washington, DC 20052, USA}
\affiliation{Institut f{\"u}r Kernphysik (Theorie),
Forschungszentrum J{\"u}lich,
D-52425 J{\"u}lich, Germany}

\date{18 July 2005 --- Revised: 19 December 2005}

\begin{abstract}
The recent high-precision data for the reaction $\gamma p\to p\eta'$ at photon
energies in the range 1.5--2.3~GeV obtained by the CLAS collaboration at the
Jefferson Laboratory have been analyzed within an extended version of the
photoproduction model developed previously by the authors based on a
relativistic meson-exchange model of hadronic interactions [Phys.\ Rev.\
C\,\textbf{69}, 065212 (2004)]. The $\eta'$ photoproduction can be described
quite well over the entire energy range of available data by considering
$S_{11}$, $P_{11}$, $P_{13}$, and $D_{13}$ resonances, in addition to the
$t$-channel mesonic currents. The observed angular distribution is due to the
interference between the $t$-channel and the nucleon $s$- and $u$-channel
resonance contributions. The $j=3/2$ resonances are required to reproduce some
of the details of the measured angular distribution. For the resonances
considered, our analysis yields mass values compatible with those advocated by
the Particle Data Group. We emphasize, however, that cross-section data alone
are unable to pin down the resonance parameters and it is shown that the beam
and/or target asymmetries impose more stringent constraints on these parameter
values. It is found that the nucleonic current is relatively small and that the
$NN\eta^\prime$ coupling constant is not expected to be much larger than 2.
\end{abstract}


\maketitle


\section{Introduction}

One of the primary interests in investigating the $\eta^\prime$ photoproduction
reaction is that it may be suited to extract information on nucleon resonances,
$N^*$, in the less explored higher $N^*$ mass region. Current knowledge of most
of the nucleon resonances is mainly due to the study of $\pi N$ scattering
and/or pion photoproduction off the nucleon. Since the $\eta'$ meson is much
heavier than a pion, $\eta'$ meson-production processes near threshold
necessarily sample a much higher resonance-mass region than the corresponding
pion production processes. They are well-suited, therefore, for investigating
high-mass resonances in low partial-wave states. Furthermore, reaction
processes such as $\eta^\prime$ photoproduction provide opportunities to study
those resonances that couple only weakly to pions, in particular, those
referred to as ``missing resonances'', which are predicted by quark models, but
not found in more traditional pion-production reactions \cite{Capstick1}.

Another special interest in $\eta ^{\prime }$ photoproduction is the
possibility to impose a more stringent constraint on its yet poorly known
coupling strength to the nucleon. This has attracted much attention in
connection with the so-called ``nucleon-spin crisis'' in polarized deep
inelastic lepton scattering \cite{EMC88}. In the zero-squared-momentum limit,
the $NN\eta ^{\prime }$ coupling constant $g_{NN\eta ^{\prime }}(q^2=0)$ is
related to the flavor-singlet axial charge $G_A$ through the flavor singlet
Goldberger--Treiman relation \cite{Shore} (see also
Refs.~\cite{Efremov,Venez1,Feldmann})
\begin{equation}
2m_N\,G_A(0) = F\,g_{NN\eta^\prime}(0) +
{\frac{F^2}{2N_F}}\,m_{\eta^\prime}^2\,g_{NNG}(0)~, \label{spinfrac}
\end{equation}
where $F$ is a renormalization-group invariant decay constant defined in
Ref.~\cite{Shore},\footnote{In the OZI limit, $F = \sqrt{2N_F}F_\pi$, where
$F_\pi$ stands for the pion decay constant.} $N_F$ is the number of flavors,
and $m_{N}$ and $m_{\eta'}$ are the nucleon and $\eta'$ masses, respectively;
$g_{NNG}$ describes the coupling of the nucleon to the gluons arising from
contributions violating the Okubo-Zweig-Iizuka (OZI) rule \cite{OZI}. The EMC
collaboration \cite{EMC88} has measured an unexpectedly small value of $G_A(0)
\approx 0.20 \pm 0.35$; a more recent analysis of the SMC collaboration
\cite{SMC97} yields a comparable value of $G_A(0) \approx 0.16 \pm 0.10$. The
first term on the right-hand side of the above equation corresponds to the
quark contribution to the ``spin'' of the proton, and the second term to the
gluon contribution \cite{Venez1,x1}. Therefore, once $g_{NN\eta'}(0)$ is known,
Eq.~(\ref{spinfrac}) may be used to extract the coupling $g_{NNG}(0)$.
Unfortunately, however, there is no direct experimental measurement of
$g_{NN\eta'}(0)$ so far. Reaction processes where the $\eta^\prime$ meson is
produced directly off a nucleon may thus offer a unique opportunity to extract
this coupling constant. Here it should be emphasized that, as has been pointed
out in Ref.~\cite{NH1}, hadronic model calculations such as the present one
cannot determine the $NN\eta'$ coupling constant in a model-independent way. At
best, we get an estimate for the range of its value at the on-shell kinematic
point, i.e., at $q^2=m_{\eta'}^2$. Assuming the usual behavior of hadronic form
factors for off-shell mesons which generally decrease for $q^2<m^2$, we expect
then that an eventually  small upper limit of $g_{NN\eta'}(q^2=m_{\eta'}^2)$
would lead to an even smaller value of $g_{NN\eta'}(0)$, which is needed in
Eq.~(\ref{spinfrac}) to extract $g_{NNG}(0)$.

The major purpose of the present work is to perform an analysis of the $\gamma
p \rightarrow p\eta^\prime$ reaction within an extended version of the
relativistic meson-exchange model of hadronic interactions as reported in
Ref.~\cite{NH1}. This analysis is motivated by the new high-precision
cross-section data obtained by the CLAS collaboration \cite{CLAS} at the
Jefferson Laboratory (JLab). The new data supersede the previous SAPHIR data
\cite{SAPHIR} analyzed in Ref.~\cite{NH1} both in absolute normalization and
angular shape. Also, the new CLAS data are much more accurate and, as such, may
reveal features that were not seen in the analysis of the SAPHIR data.

The present paper is organized as follows. In Sec.~II the extension of our
model~\cite{NH1} for $\gamma p \rightarrow p\eta^\prime$ is given. The results
of the corresponding model calculations are presented in Sec.~III. Section~IV
contains a summary with our conclusions. Some technical details of the present
model are given in the Appendix.


\section{Formalism}

\begin{figure}[t!]
\includegraphics[width=.8\columnwidth,angle=0,clip]{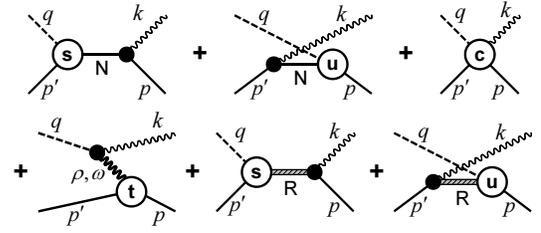}
\caption{\label{fig:diagram}%
Diagrams contributing to $\gamma p \to \eta' p$. Time proceeds from right to
left. The intermediate baryon states are denoted \textsf{N} for the nucleon,
and \textsf{R} for the $S_{11}$, $P_{11}$, $P_{13}$, and $D_{13}$ resonances.
The intermediate mesons in the $t$-channel are $\rho$ and $\omega$. The
external legs are labeled by the four-momenta of the respective particles and
the labels \textsf{s}, \textsf{u}, and \textsf{t} of the hadronic vertices
correspond to the off-shell Mandelstam variables of the respective intermediate
particles. The three diagrams in the lower part of the diagram are transverse
individually; the three diagrams in the upper part are made gauge-invariant by
an appropriate choice for the contact current depicted in the top-right
diagram. The nucleonic current (nuc) referred to in the text corresponds to the
top line of diagrams; the meson-exchange current (mec) and resonance current
contributions correspond, respectively, to the leftmost diagram and the two
diagrams on the right of the bottom line of diagrams.}
\end{figure}

The dynamical content of the present $\eta'$ photoproduction calculation is
summarized by the graphs of Fig.~\ref{fig:diagram} where we employ form factors
at the vertices to account for the hadronic structure. The gauge invariance of
this production current is ensured by a phenomenological contact current,
according to the prescription of Refs.~\cite{hh97g,hhtree98,dw2}. This contact
term provides a rough phenomenological description of the final-state
interaction which is not treated explicitly here. The basic details of the
present approach are the same as in our previous paper~\cite{NH1} and we will
not repeat them here. There are, however, a few improvements and those will be
discussed here.

\subsection{Spin-3/2 resonances}

The present fits also require the inclusion of spin-3/2 resonances, denoted
generically by $N^*$. The Lagrangian for the hadronic $N N^*\eta' $ interaction
is given by
\begin{equation}
\mathcal{L}^{(\pm)}_{N N^*\eta'}= \frac{g_{NN^*\eta'}}{m_{\eta'}}
{\bar{N}^{*\mu}} \Theta_{\mu\nu}(z) \Gamma^{(\pm)} N
\partial^\nu\,\eta' +\hc~,
 \label{eq:H32}
\end{equation}
where $N^{*\mu}$, $N$, and $\eta'$ are the resonance, nucleon, and meson
fields, respectively, and
\begin{equation}
\Gamma^{(+)}=\gamma_5 \quad\text{and}\quad \Gamma^{(-)}=1
\end{equation}
pertain to positive- and negative-parity resonances, respectively. For the
coupling tensor,
$\Theta_{\mu\nu}=g_{\mu\nu}-(z+\frac{1}{2})\gamma_\mu\gamma_\nu$, we take
$z=-\frac{1}{2}$ for the off-shell parameter for simplicity.\footnote{%
We have also explored how the fits changed upon varying the off-shell parameter
$z$ in Eqs.~(\ref{eq:H32}) and (\ref{eq:E32}), since spin-3/2 resonances also
play a relevant role in reproducing the data quantitatively, as discussed in
Sec.~\ref{sec:results}. However, we didn't observe any significant changes due
to this parameter and we feel justified, therefore, to keep this parameter at
$z=-\frac{1}{2}$ for simplicity.} The Lagrangian for the electromagnetic
transition current reads
\begin{align}
\mathcal{L}^{(\pm)}_{NN^*\gamma}&=-ie\frac{g_{1NN^*\gamma}}{m_{N^*}}
{\bar{N}^{*\beta}} \Theta_{\beta\mu} \Gamma^{(\pm)}\gamma_\nu N F^{\mu\nu}
\nonumber\\
&\quad\mbox{} -e\frac{g_{2NN^*\gamma}}{2m^2_{N^*}} \left( \partial_\nu
\bar{N}^{*\beta} \Theta_{\beta\mu} \Gamma^{(\pm)} N \right) F^{\mu\nu} +\hc~,
\label{eq:E32}
\end{align}
where $F^{\mu\nu}=\partial^\mu A^\nu-\partial^\nu A^\mu$ is the
electromagnetic field-strength tensor (with $A^\mu$ being the vector
potential).

\subsection{Energy-dependent resonance widths}\label{sec:widths}

For the present application, we have adapted our formalism to accommodate
energy-dependent resonance widths with the appropriate threshold behavior.

For a spin-1/2 resonance propagator, we use the ansatz
\begin{equation}
S_{1/2}(p) = \frac{1}{\fs{p}-m_R
+\frac{i}{2}\Gamma}=\frac{\fs{p}+m_R}{p^2-m_R^2+\frac{i}{2}(\fs{p}+m_R)\Gamma}~,
\label{eq:spin1}
\end{equation}
where $m_R$ is the mass of the resonance with four-momentum $p$. $\Gamma$ is
the width function whose functional behavior will be given below.

For spin-3/2, the resonant propagator reads in a schematic matrix notation
\begin{equation}
S_{3/2}(p)=\left[(\fs{p}-m_R)g-i\frac{\Delta}{2}\Gamma\right]^{-1}\Delta~.
\label{eq:spin3}
\end{equation}
All indices are suppressed here, i.e., $g$ is the metric tensor and $\Delta$ is
the Rarita--Schwinger tensor written in full detail as
\begin{equation}
\Delta^{\mu\nu}_{\beta\alpha}=
-g^{\mu\nu}\delta_{\beta\alpha}+\frac{1}{3}\gamma^\mu_{\beta\varepsilon}\gamma^\nu_{\varepsilon\alpha}
        + \frac{2p^\mu p^\nu}{3m_R^2}\delta_{\beta\alpha}
                 +\frac{\gamma^\mu_{\beta\alpha} p^\nu-p^\mu\gamma^\nu_{\beta\alpha}}{3m_R}~,
\label{eq:RStensor}
\end{equation}
where $\beta$, $\alpha$, and $\varepsilon$ enumerate the four indices of the
$\gamma$-matrix components (summation over $\varepsilon$ is implied). The
inversion in (\ref{eq:spin3}) is to be understood on the full 16-dimensional
space of the four Lorentz indices and the four components of the gamma
matrices. The motivation for the ansatz (\ref{eq:spin3}) and the technical
details how to perform this inversion is given in the Appendix.

In both cases, we write the width $\Gamma$ as a function of $W=\sqrt{s}$
according to
\begin{equation}
\Gamma(W) = \Gamma_R\left[\sum_{i=1}^N \beta_i \hat{\Gamma}_i(W) +
\sum_{j=1}^{N_\gamma} \gamma_j \hat{\Gamma}_{\gamma_j}(W)\right]~,
\end{equation}
where the sums over $i$ and $j$ respectively account for decays of the
resonance into $N$ two- or three-hadron channels and into $N_\gamma$ radiative
decay channels. The total static resonance width is denoted by $\Gamma_R$ and
the numerical factors $\beta_i$ and $\gamma_j$ (with $0\le \beta_i,\gamma_j\le
1$) describe the branching ratios into the various decay channels, i.e.,
\begin{equation}
\sum_{i=1}^N \beta_i +\sum_{j=1}^{N_\gamma}\gamma_j =1~.
\end{equation}
Similar to Refs.~\cite{walker,arndt90,lvov97,drechsel}, we parameterize the
width functions $\hat{\Gamma}_i$ and $\hat{\Gamma}_{\gamma_j}$ (which are both
normalized to unity at $W=m_R$) to provide the correct respective threshold
behaviors.

For the decay of the resonance into two hadronic fragments with masses
${m_i}_1$ and ${m_i}_2$, the hadronic width functions $\hat{\Gamma}_i$ are
taken as
\begin{equation}
\hat{\Gamma}_i(W)=\left(\frac{\mathrm{q}_i}{\mathrm{q}_{iR}}\right)^{2L+1}
\left(\frac{\lambda_i^2+\mathrm{q}^2_{iR}}{\lambda_i^2+\mathrm{q}^2_{i}}\right)^L
D_i(W)
\end{equation}
for $W>{m_i}_1+{m_i}_2$, and zero otherwise. $L$ denotes the partial wave in
which the resonance is found and the momentum $\mathrm{q}_i$ is the magnitude
of the center-of-momentum three-momentum of the two fragments, i.e.,
\begin{equation}
\mathrm{q}_i(W) =
\frac{\sqrt{\left[W^2-({m_i}_1+{m_i}_2)^2\right]\left[W^2-({m_i}_1-{m_i}_2)^2\right]}}{2W}
\label{eq:qW}
\end{equation}
and $\mathrm{q}_{iR}=\mathrm{q}_{i}(m_R)$. For the decay of the resonance into
one baryon and two mesons (for example, $N\pi\pi$), we use
\begin{equation}
\hat{\Gamma}_i(W)=\left(\frac{\mathrm{q}_i}{\mathrm{q}_{iR}}\right)^{2L+4}
\left(\frac{\lambda_i^2+\mathrm{q}^2_{iR}}{\lambda_i^2+\mathrm{q}^2_{i}}\right)^{L+2}
D_i(W)~,
 \label{eq:width2}
\end{equation}
where ${m_i}_2$ in (\ref{eq:qW}) needs to be replaced by the sum of the two
meson masses for this case, and ${m_i}_1$ is the baryon mass. In principle, the
factor
\begin{equation}
D_i(W)=\left(\frac{m_R}{W}\right)^{n_i}~, \quad\text{with}\quad n_i
\ge 0~,
 \label{eq:width3}
\end{equation}
allows for a modification of the asymptotic behavior of $\hat{\Gamma}_i(W)$,
however, we use $n_i=0$ throughout for simplicity. The parameter $\lambda_i$ is
an inverse range parameter; since we found very little sensitivity to varying
this parameter (within reasonable ranges), we kept it fixed at
$\lambda_i=1\,\text{fm}^{-1}$ for all channels.

The width function $\hat{\Gamma}_{\gamma_j}$ for the decay into a hadron with
mass $m_j$ and a photon with three-momentum $\mathrm{k}_j$ is taken as
\begin{equation}
\hat{\Gamma}_{\gamma_j}(W)=\left(\frac{\mathrm{k}_j}{\mathrm{k}_{jR}}\right)^{2L+2}
\left(\frac{\lambda_{\gamma_j}^2+\mathrm{k}^2_{jR}}{\lambda_{\gamma_j}^2+\mathrm{k}^2_{j}}\right)^{L+1}
D_{\gamma_j}(W)~,
 \label{eq:widthg}
\end{equation}
where
\begin{equation}
\mathrm{k}_{j}(W)=\frac{W^2-m_j^2}{2W}
\end{equation}
for $W>m_j$, and zero otherwise, and $\mathrm{k}_{jR}=\mathrm{k}_{j}(m_R)$. As
in the hadronic case, the asymptotic damping function is given by
\begin{equation}
D_{\gamma_j}(W) = \left(\frac{m_R}{W}\right)^{r_j}~,
\quad\text{with}\quad r_j \ge 0~.
\end{equation}
Again, for simplicity, we employ $r_j=0$ throughout. In practice, for the
present case, the photon decay channels are negligibly small and play no role
for the total width. The corresponding branching ratio $\gamma_{N\gamma}$ for
the $N\gamma$ channel is only needed to extract the value of the $N\eta'$
branching ratio $\beta_{N\eta'}$ (see below).

\section{Results and Discussion}\label{sec:results}

Before we discuss the details of our results, some general remarks are in
order. The basic strategy of our model approach is to start with the nucleon
plus meson-exchange currents and add the resonances one by one as needed in the
fitting procedure until one achieves a reasonable fit of the new $\eta'$
photoproduction data obtained by the CLAS collaboration~\cite{CLAS}. We allow
for both spin-1/2 and \mbox{-3/2} resonances in our model. Our quantitative
criterion for a reasonable fit was to discard all fits with a $\chi^2$ per data
point of $\chi^2/N > 1.3$, which is supported by the fact that fits with
$\chi^2/N$ much larger than 1.3 are noticeably of inferior fit quality even for
the naked eye. Under this criterion, we found that one needs at least four
resonances in order to obtain a reasonable fit in the present approach. We
find, in particular, that, in addition to spin-1/2 resonances, spin-3/2
resonances are necessary to achieve acceptable fits.
In this respect, we emphasize that the SAPHIR data~\cite{SAPHIR}
analyzed in Ref.~\cite{NH1} have rather large error bars. While not entirely
incompatible with the new high-precision CLAS data \cite{CLAS}, they clearly
are less constraining than the CLAS data, which may explain why there was no
need for spin-3/2 resonances in our previous work.

As we have pointed out in Ref.~\cite{NH1}, the cross-section data alone are
unable to pin down the model parameters and, therefore, one finds different
sets of parameters which fit the data equally well. Note that this is not due
to the uncertainties in the data, but simply because, intrinsically, the cross
sections do not impose enough stringent constraints on the fit. In particular,
for each resonance, the resulting fitted mass value depends to a certain extent
on its starting value in the fitting procedure. The starting (resonance) mass
values we consider here generally are around those advocated by the Particle
Data Group (PDG)~\cite{PDG}.

In the present work, in the case of those resonances that can be identified
with known PDG resonances, we have taken into account only the corresponding
dominant branching ratios $\beta_i$ from the PDG for hadronic decays when this
information is available (and we ignored the fact that some of the quoted
branching ratios are subject to large uncertainties). If no information is
available, we consider only the $N\pi$ partial decay, with the corresponding
branching ratio $\beta_{N\pi}$ as a free fit parameter. Apart from these
branching ratios, we also consider the $N\eta'$ branching ratio
$\beta_{N\eta'}$ which is calculated from the product of the coupling constants
$g_{NN^*\eta'}g_{NN^*\gamma}$ in conjunction with the assumed branching ratio
$\gamma_{N\gamma}$ for the radiative decay. In the following tables, therefore,
$\beta_{N\eta'}$ is not an independent fit parameter, but rather a parameter
extracted from the fitted values of the product $g_{NN^*\eta'}g_{NN^*\gamma}$.

One might expect that the way in which the energy dependence is implemented in
the resonance width in the present work [cf.\
Eqs.~(\ref{eq:width2})--(\ref{eq:widthg})] may introduce a considerable
uncertainty in the final results. However, we find that the cross sections are
not very sensitive to our assumptions in this respect. In fact, we also re-ran
some of the cross-section fits discussed below using step functions for the
widths that switch on the full partial widths at the corresponding thresholds
without any smooth energy dependence and we found that the parameter sets
obtained in this way were fairly close to the ones reported here. For spin
observables, however, this insensitivity does not hold true. In particular, the
beam and target asymmetries are rather sensitive to how the energy dependence
of the width is treated and one must be careful then when confronting model
predictions with the data when the latter should become available.

\begin{figure}[t!]
\includegraphics[width=\columnwidth,angle=0,clip]{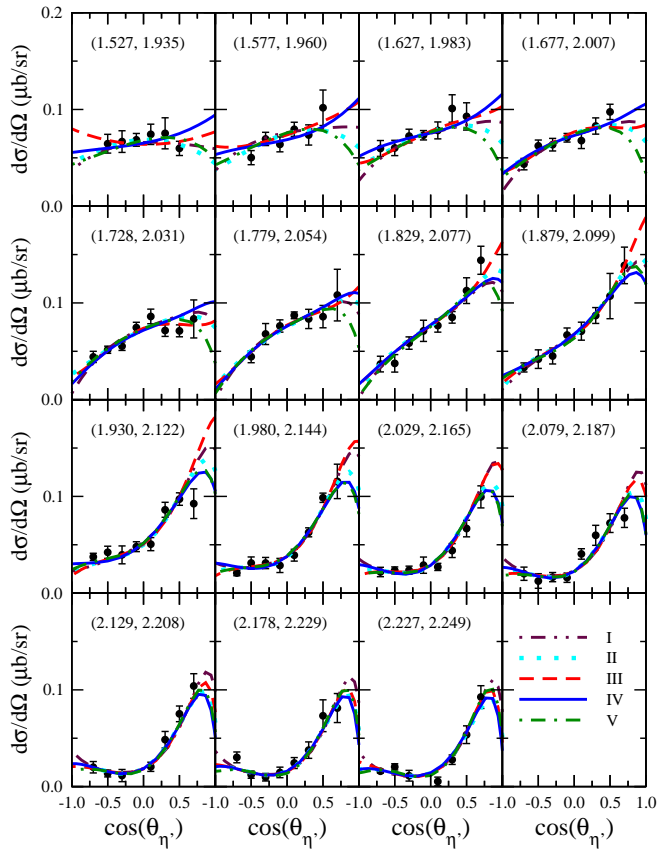}
\caption{\label{fig:Rall}%
(Color online) Differential cross section for $\gamma p\to p \eta'$ according
to the mechanisms shown in Fig.~\ref{fig:diagram} as a function of the $\eta'$
emission angle $\theta_{\eta'}$ in the center-of-momentum frame of the system.
As indicated in the legend, the curves correspond to the fit results of
Tables~\ref{tbl:R1275} (dash-double-dotted), \ref{tbl:R1308} (dotted lines),
\ref{tbl:R1515} (dashed),
\ref{tbl:R1428} (solid), and \ref{tbl:R1447} (dash-dotted). The numbers
($T\gamma , W$) in parentheses are the incident photon energy $T_\gamma$ and
the corresponding $s$-channel energy $W=\sqrt{s}$, respectively, in GeV. The
data are from Ref.~\cite{CLAS}. }
\end{figure}
%

We now turn to the discussion of the details of our analysis. We emphasize that
the results shown here do not necessarily have the lowest $\chi^2$. Rather,
they are sample fit results that illustrate the different dynamical features
one may obtain considering only the currently available data in the analysis
within the fit-quality criteria mentioned above.


\begin{table}[t!]
\begin{center}
\caption{Model parameters fitted to the $\gamma p \to  \eta' p$. (See text, and
also Ref.~\cite{NH1}, for explanations of parameters.) Values in boldface are
not fitted. The branching ratios $\gamma_{N\gamma}$ are assumptions made to
extract $\beta_{N\eta^\prime}$ (for the total width, however, the
$\gamma_{N\gamma}$ values are too small to be relevant).
 The starting values for
fitting all resonance masses were chosen here within the energy range covered
by the data set. $\chi^2/N=1.19$. }
\begin{tabular}{l@{\qquad}r@{\qquad}r@{\qquad}r}
\hline\hline
Nucleonic current: & & &  \\
$g_{NN\eta^\prime}$                      & 0.43           &                &                   \\
$\lambda$                                & \textbf{0.0}   &                &                   \\
$\Lambda_N$~(MeV)                        & \textbf{1200}  &                &                   \\
\hline
Mesonic current: & & &   \\
$g_{\eta^\prime\rho\gamma}$              & \textbf{1.25}  &                &                   \\
$g_{\eta^\prime\omega\gamma}$            & \textbf{0.44}  &                &                   \\
$\Lambda_v$ (MeV)                        &  1275          &                &                   \\
\hline
$N^*=S_{11}$ current: & & &   \\
$m_{N^*}$ (MeV)                          & 1958           &                &                   \\
$g_{NN^*\gamma}\,g_{NN^*\eta^\prime}$    &  0.25          &                &                   \\
$\lambda$                                &  1.00          &                &                   \\
$\Lambda_{N^*}$ (MeV)                    & \textbf{1200}  &                &                   \\
$\Gamma_{N^*} $ (MeV)                    &  139           &                &                   \\
$\gamma_{N\gamma}  $                     & \textbf{0.002} &                &                   \\
$\beta_{N\pi}  $                         & 0.50           &                &                   \\
$\beta_{N\eta^\prime} $                  & 0.50           &                &                   \\
\hline
$N^*=P_{11}$ current: & & &  \\
$m_{N^*}$ (MeV)                          & 2104           &                &                   \\
$g_{NN^*\gamma}\,g_{NN^*\eta^\prime}$    & 0.80           &                &                   \\
$\lambda$                                &  1.00          &                &                   \\
$\Lambda_{N^*}$ (MeV)                    & \textbf{1200}  &                &                   \\
$\Gamma_{N^*} $ (MeV)                    &  136           &                &                   \\
$\gamma_{N\gamma}  $                     & \textbf{0.002} &                &                   \\
$\beta_{N\pi}  $                         & 0.36           &                &                   \\
$\beta_{N\eta^\prime} $                  & 0.64           &                &                   \\
\hline
$N^*=P_{13}$ current: & & &  \\
$m_{N^*}$ (MeV)                          &  1885          &                &                   \\
$g_{1NN^*\gamma}\,g_{NN^*\eta^\prime}$   & 0.01           &                &                   \\
$g_{2NN^*\gamma}\,g_{NN^*\eta^\prime}$   & 0.17           &                &                   \\
$\Lambda_{N^*}$ (MeV)                    & \textbf{1200}  &                &                   \\
$\Gamma_{N^*} $ (MeV)                    &  59            &                &                   \\
$\beta_{N\pi}  $                         & \textbf{0.6}   &                &                   \\
$\beta_{N\omega}$                        & \textbf{0.4}   &                &                   \\
\hline
$N^*=D_{13}$ current: & & & \\
$m_{N^*}$ (MeV)                          & 1823           &                &                   \\
$g_{1NN^*\gamma}\,g_{NN^*\eta^\prime}$   & 0.47           &                &                   \\
$g_{2NN^*\gamma}\,g_{NN^*\eta^\prime}$   & -0.65          &                &                   \\
$\Gamma_{N^*} $ (MeV)                    & 450            &                &                   \\
$\gamma_{N\gamma}  $                     & \textbf{0.002} &                &                   \\
$\beta_{N\pi}  $                         & 1.00           &                &                   \\
\hline\hline
\end{tabular}
\label{tbl:R1275}
\end{center}
\vspace{-8ex}
\end{table}
%
\begin{figure}[t!]
\includegraphics[width=\columnwidth,angle=0,clip]{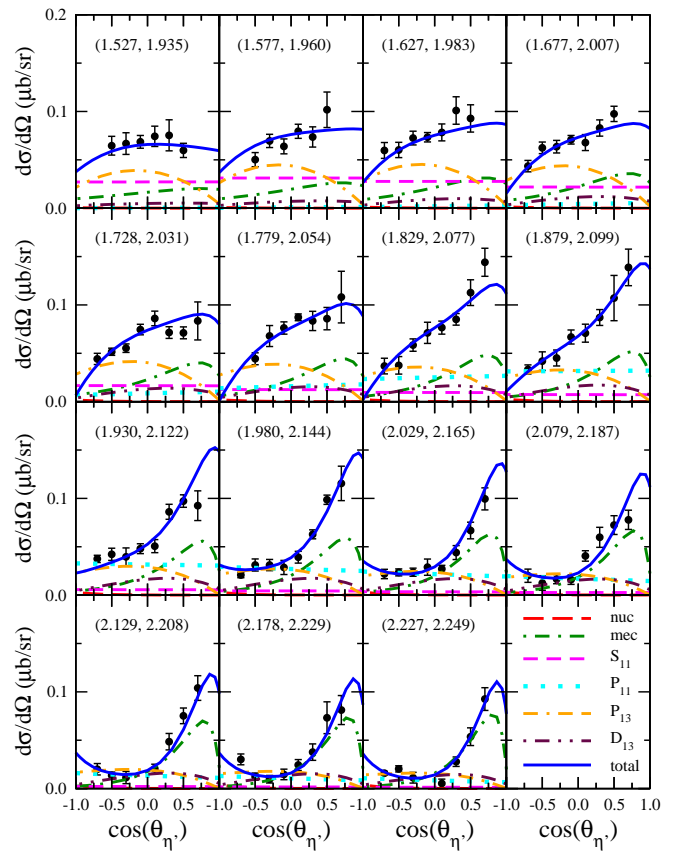}
\caption{\label{fig:R1275}%
(Color online) Differential cross sections and the dynamical content of the
present model corresponding to the fit result of Table~\ref{tbl:R1275}. The
dash-dotted curves correspond to the mesonic current contribution; the dashed
curves to the $S_{11}$ resonance current and the dotted curves to the $P_{11}$
resonance. The dot--double-dashed curves correspond to the $P_{13}$ resonance
current while the dash--double-dotted curves show the $D_{13}$ resonance
contribution. The solid curves correspond to the total current.
The nucleonic current contribution (long-dashed curves)
is very small and cannot be seen on the present scale.}
\end{figure}
%

\begin{table}[t!]
\begin{center}
\caption{Same as Table.~\ref{tbl:R1275}. An additional $D_{13}$ resonance was
included here to see whether this would improve the fit quality.
$\chi^2/N=1.04$. }
\begin{tabular}{l@{\qquad}r@{\qquad}r@{\qquad}r}
\hline\hline
Nucleonic current: & & &  \\
$g_{NN\eta^\prime}$                      & 0.25           &                &                   \\
$\lambda$                                & \textbf{0.0}   &                &                   \\
$\Lambda_N$~(MeV)                        & \textbf{1200}  &                &                   \\
\hline
Mesonic current: & & &   \\
$g_{\eta^\prime\rho\gamma}$              & \textbf{1.25}  &                &                   \\
$g_{\eta^\prime\omega\gamma}$            & \textbf{0.44}  &                &                   \\
$\Lambda_v$ (MeV)                        &  1308          &                &                   \\
\hline
$N^*=S_{11}$ current: & & &   \\
$m_{N^*}$ (MeV)                          & 1925           &                &                   \\
$g_{NN^*\gamma}\,g_{NN^*\eta^\prime}$    &  0.08          &                &                   \\
$\lambda$                                &  0.58          &                &                   \\
$\Lambda_{N^*}$ (MeV)                    & \textbf{1200}  &                &                   \\
$\Gamma_{N^*} $ (MeV)                    &  40            &                &                   \\
$\gamma_{N\gamma}  $                     & \textbf{0.002} &                &                   \\
$\beta_{N\pi}  $                         & 0.56           &                &                   \\
$\beta_{N\eta^\prime} $                  & 0.44           &                &                   \\
\hline
$N^*=P_{11}$ current: & & &  \\
$m_{N^*}$ (MeV)                          & 1991           &                &                   \\
$g_{NN^*\gamma}\,g_{NN^*\eta^\prime}$    & -1.69          &                &                   \\
$\lambda$                                &  0.09          &                &                   \\
$\Lambda_{N^*}$ (MeV)                    & \textbf{1200}  &                &                   \\
$\Gamma_{N^*} $ (MeV)                    &  158           &                &                   \\
$\gamma_{N\gamma}  $                     & \textbf{0.002} &                &                   \\
$\beta_{N\pi}  $                         & 0.42           &                &                   \\
$\beta_{N\eta^\prime} $                  & 0.58           &                &                   \\
\hline
$N^*=P_{13}$ current: & & &  \\
$m_{N^*}$ (MeV)                          &  1907          &                &                   \\
$g_{1NN^*\gamma}\,g_{NN^*\eta^\prime}$   & -0.06          &                &                   \\
$g_{2NN^*\gamma}\,g_{NN^*\eta^\prime}$   & -0.09          &                &                   \\
$\Lambda_{N^*}$ (MeV)                    & \textbf{1200}  &                &                   \\
$\Gamma_{N^*} $ (MeV)                    &  123           &                &                   \\
$\gamma_{N\gamma}  $                     & \textbf{0.002} &                &                   \\
$\beta_{N\pi}  $                         & 0.60           &                &                   \\
$\beta_{N\omega}$                        & \textbf{0.4}   &                &                   \\
$\beta_{N\eta^\prime} $                  & 0.00           &                &                   \\
\hline
$N^*=D_{13}$ current: & & & \\
$m_{N^*}$ (MeV)                          & 1825           & 2084           &                   \\
$g_{1NN^*\gamma}\,g_{NN^*\eta^\prime}$   & -1.17          & -0.21          &                   \\
$g_{2NN^*\gamma}\,g_{NN^*\eta^\prime}$   &  0.53          & 0.19           &                   \\
$\Lambda_{N^*}$ (MeV)                    & \textbf{1200}  & \textbf{1200}  &                   \\
$\Gamma_{N^*} $ (MeV)                    & 55             &  108           &                   \\
$\gamma_{N\gamma}  $                     & \textbf{0.002} & \textbf{0.002} &                   \\
$\beta_{N\pi}  $                         & 1.00           & 0.54           &                   \\
$\beta_{N\eta^\prime} $                  & 0.00           &  0.46          &                   \\
\hline\hline
\end{tabular}
\label{tbl:R1308}
\end{center}
\vspace{-4ex}
\end{table}
%
\begin{figure}[t!]
\includegraphics[width=\columnwidth,angle=0,clip]{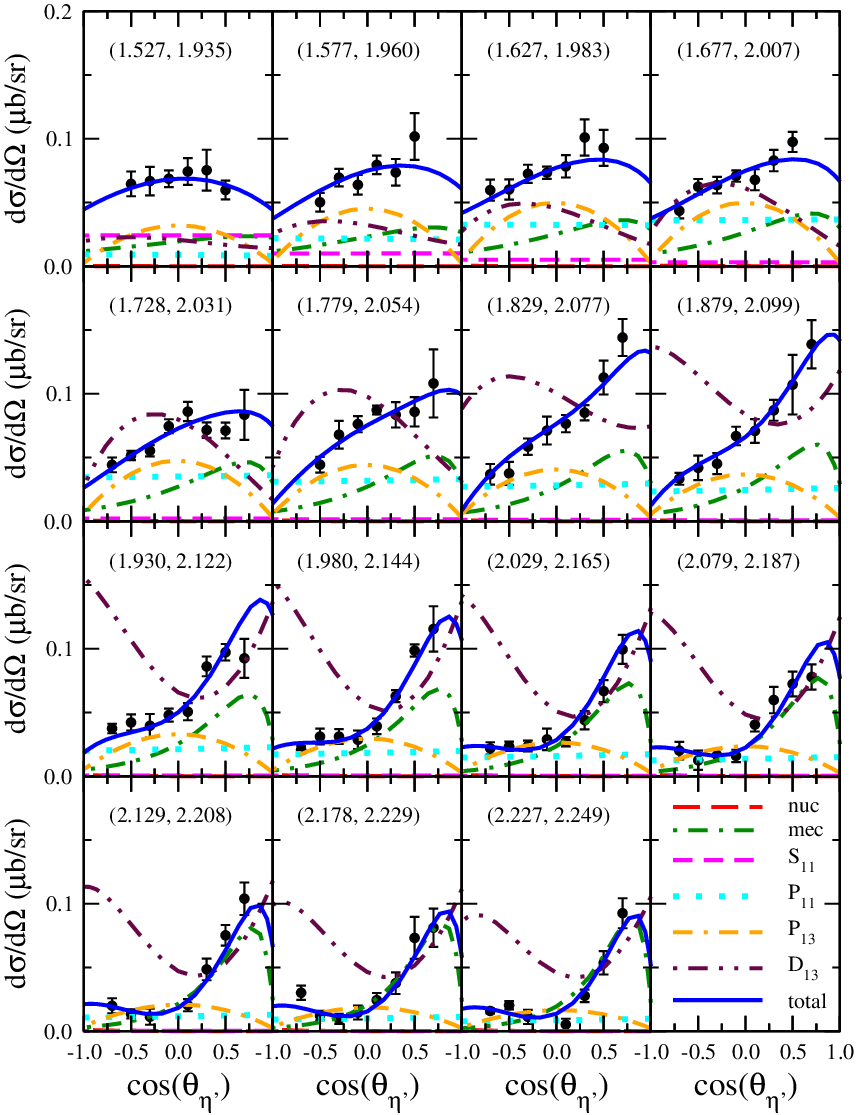}
\caption{\label{fig:R1308}%
(Color online) Same as in Fig.~\ref{fig:R1275} for the fit result of
Table~\ref{tbl:R1308}.
The nucleonic current contribution (long-dashed curves)
is very small and cannot be seen on the present scale.}
\end{figure}
%

\begin{table}[t!]
\begin{center}
\caption{Same as Table~\ref{tbl:R1308}. More resonances were added here to see
whether this would further improve the fit.
 $\chi^2/N=1.04$.}
\begin{tabular}{l@{\qquad}r@{\qquad}r@{\qquad}r@{\qquad}r}
\hline\hline
Nucleonic current: & & & & \\
$g_{NN\eta^\prime}$            & 1.33               &                   &                 &        \\
$\lambda$                      & \textbf{0.0}       &                   &                 &        \\
$\Lambda_N$~(MeV)              &  \textbf{1200}     &                   &                 &        \\
\hline
Mesonic current: & & & &  \\
$g_{\eta^\prime\rho\gamma}$    & \textbf{1.25}      &                   &                 &       \\
$g_{\eta^\prime\omega\gamma}$  & \textbf{0.44}      &                   &                 &       \\
$\Lambda_v$ (MeV)              &  1515              &                   &                 &       \\
\hline
$N^*=S_{11}$ current: & & & &  \\
$m_{N^*}$ (MeV)                       & 1539            &  1670          & 2025           &     \\
$g_{NN^*\gamma}\,g_{NN^*\eta^\prime}$ & -6.48           &  1.10          & 0.03           &     \\
$\lambda$                             & 0.78            &  0.93          &  0.07          &     \\
$\Lambda_{N^*}$ (MeV)                 & \textbf{1200}   &\textbf{1200}   &\textbf{1200}   &     \\
$\Gamma_{N^*} $ (MeV)                 & 138             &  79            & 79             &     \\
$\gamma_{N\gamma}  $                  &                 &                & \bf{0.001}     &     \\
$\beta_{N\pi}  $                      & \textbf{0.5}    &\textbf{0.9}    & 0.96           &     \\
$\beta_{N\eta}  $                     & \textbf{0.5}    & \textbf{0.1}   &                &     \\
$\beta_{N\eta^\prime} $               &                 &                & 0.04           &     \\
\hline
$N^*=P_{11}$ current: & & & & \\
$m_{N^*}$ (MeV)                       &  1718           &  2099          &  2406          &    \\
$g_{NN^*\gamma}\,g_{NN^*\eta^\prime}$ & 1.45            & -0.90          & -0.27          &    \\
$\lambda$                             & 1.00            & 0.78           & 0.71           &    \\
$\Lambda_{N^*}$ (MeV)                 & \textbf{1200}   & \textbf{1200}  & \text bf{1200} &    \\
$\Gamma_{N^*} $ (MeV)                 &  89             &  172           &  82            &    \\
$\gamma_{N\gamma}  $                  &                 & \textbf{0.002} & \textbf{0.002} &    \\
$\beta_{N\pi}  $                      & \textbf{0.15}   & 0.51           &  0.00          &    \\
$\beta_{N\pi\pi}  $                   & \textbf{0.85}   &                &                &    \\
$\beta_{N\eta^\prime} $               &                 & 0.49           &  1.00          &    \\
\hline
$N^*=P_{13}$ current: & & & &  \\
$m_{N^*}$ (MeV)                        &  1943          &                &                &    \\
$g_{1NN^*\gamma}\,g_{NN^*\eta^\prime}$ & 0.06           &                &                &    \\
$g_{2NN^*\gamma}\,g_{NN^*\eta^\prime}$ & -0.13          &                &                &    \\
$\Lambda_{N^*}$ (MeV)                  & \textbf{1200}  &                &                &    \\
$\Gamma_{N^*} $ (MeV)                  &  109           &                &                &     \\
$\gamma_{N\gamma}  $                   & \textbf{0.002} &                &                &     \\
$\beta_{N\pi}  $                       & 0.59           &                &                &     \\
$\beta_{N\omega}$                      & \textbf{0.4}   &                &                &     \\
$\beta_{N\eta^\prime} $                & 0.01           &                &                &     \\
\hline
$N^*=D_{13}$ current: & & & & \\
$m_{N^*}$ (MeV)                        & 1782           & 2085           &                &    \\
$g_{1NN^*\gamma}\,g_{NN^*\eta^\prime}$ & -0.17          & -0.01          &                &    \\
$g_{2NN^*\gamma}\,g_{NN^*\eta^\prime}$ & -0.24          & 0.10           &                &    \\
$\Lambda_{N^*}$ (MeV)                  & \textbf{1200}  & \textbf{1200}  &                &     \\
$\Gamma_{N^*} $ (MeV)                  & 152            &  141           &                &     \\
$\gamma_{N\gamma}  $                   &                &\textbf{0.001}  &                &     \\
$\beta_{N\pi}  $                       & \textbf{0.1}   & 0.97           &                &     \\
$\beta_{N\pi\pi}  $                    & \textbf{0.9}   &                &                &     \\
$\beta_{N\eta^\prime} $                &                & 0.03           &                &     \\
\hline\hline
\end{tabular}
\label{tbl:R1515}
\end{center}
\vspace{-8ex}
\end{table}
%
\begin{figure}[t!]
\includegraphics[width=\columnwidth,angle=0,clip]{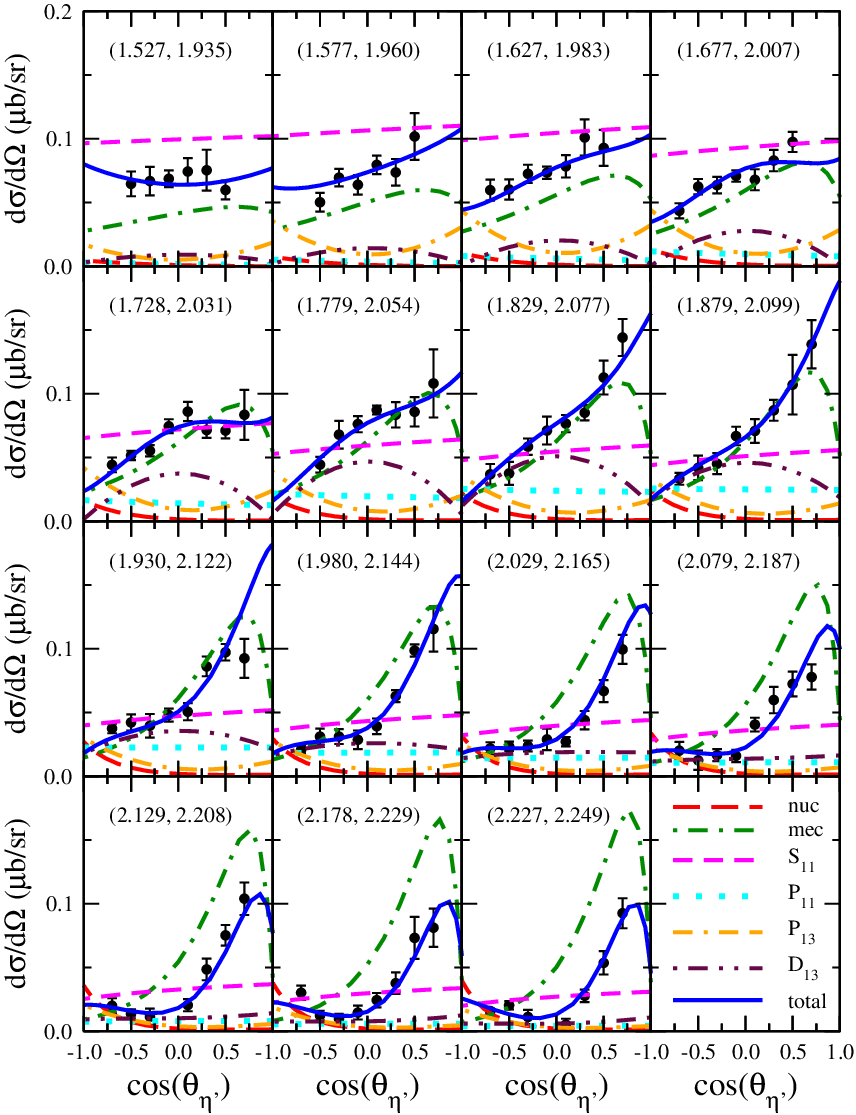}
\caption{\label{fig:R1515}%
(Color online) Same as in Fig.~\ref{fig:R1275} for the fit result of
Table~\ref{tbl:R1515}. }
\end{figure}
%

\begin{table}[t!]
\begin{center}
\caption{Same as Table~\ref{tbl:R1275}. No $P_{13}$ resonance was allowed here.
$\chi^2/N=1.10$. }
\begin{tabular}{l@{\qquad}r@{\qquad}r@{\qquad}r@{\qquad}r}
\hline\hline
Nucleonic current: & & & & \\
$g_{NN\eta^\prime}$                      & 0.002          &                &                  &                  \\
$\lambda$                                & \textbf{0.0}   &                &                  &                  \\
$\Lambda_N$~(MeV)                        & \textbf{1200}  &                &                  &                  \\
\hline
Mesonic current: & & & &  \\
$g_{\eta^\prime\rho\gamma}$              & \textbf{1.25}  &                &                  &                  \\
$g_{\eta^\prime\omega\gamma}$            & \textbf{0.44}  &                &                  &                  \\
$\Lambda_v$ (MeV)                        &  1428          &                &                  &                  \\
\hline
$N^*=S_{11}$ current: & & & &  \\
$m_{N^*}$ (MeV)                          & 1542           & 1848           &                  &                  \\
$g_{NN^*\gamma}\,g_{NN^*\eta^\prime}$    & -10.33         &  2.12          &                  &                  \\
$\lambda$                                & 1.00           & 1.00           &                  &                  \\
$\Lambda_{N^*}$ (MeV)                    & \textbf{1200}  & \textbf{1200}  &                  &                  \\
$\Gamma_{N^*} $ (MeV)                    & 233            & 164            &                  &                  \\
$\beta_{N\pi}  $                         & \textbf{0.5}   & 1.00           &                  &                  \\
$\beta_{N\eta}  $                        & \textbf{0.5}   &                &                  &                  \\
\hline
$N^*=P_{11}$ current: & & & & \\
$m_{N^*}$ (MeV)                          &  1710          &  1996          &                  &                  \\
$g_{NN^*\gamma}\,g_{NN^*\eta^\prime}$    &  4.34          & -1.37          &                  &                  \\
$\lambda$                                & 1.00           &  0.13          &                  &                  \\
$\Lambda_{N^*}$ (MeV)                    & \textbf{1200}  & \textbf{1200}  &                  &                  \\
$\Gamma_{N^*} $ (MeV)                    &  39            &  118           &                  &                  \\
$\gamma_{N\gamma}  $                     &                & \textbf{0.002} &                  &                  \\
$\beta_{N\pi}  $                         & \textbf{0.15}  & 0.26           &                  &                  \\
$\beta_{N\pi\pi}  $                      & \textbf{0.85}  &                &                  &                  \\
$\beta_{N\eta^\prime} $                  &                & 0.74           &                  &                  \\
\hline
$N^*=D_{13}$ current: & & & & \\
$m_{N^*}$ (MeV)                          & 1756           & 2087           &                  &                  \\
$g_{1NN^*\gamma}\,g_{NN^*\eta^\prime}$   & -0.67          & -0.08          &                  &                  \\
$g_{2NN^*\gamma}\,g_{NN^*\eta^\prime}$   &  0.02          & 0.15           &                  &                  \\
$\Lambda_{N^*}$ (MeV)                    & \textbf{1200}  & \textbf{1200}  &                  &                  \\
$\Gamma_{N^*} $ (MeV)                    &  48            &  134           &                  &                  \\
$\gamma_{N\gamma}  $                     &                & \textbf{0.002} &                  &                  \\
$\beta_{N\pi}  $                         & 1.00           & 0.88           &                  &                  \\
$\beta_{N\eta^\prime} $                  &                & 0.12           &                  &                  \\
\hline\hline
\end{tabular}
\label{tbl:R1428}
\end{center}
\vspace{-4ex}
\end{table}
%
\begin{figure}[t!]
\includegraphics[width=\columnwidth,angle=0,clip]{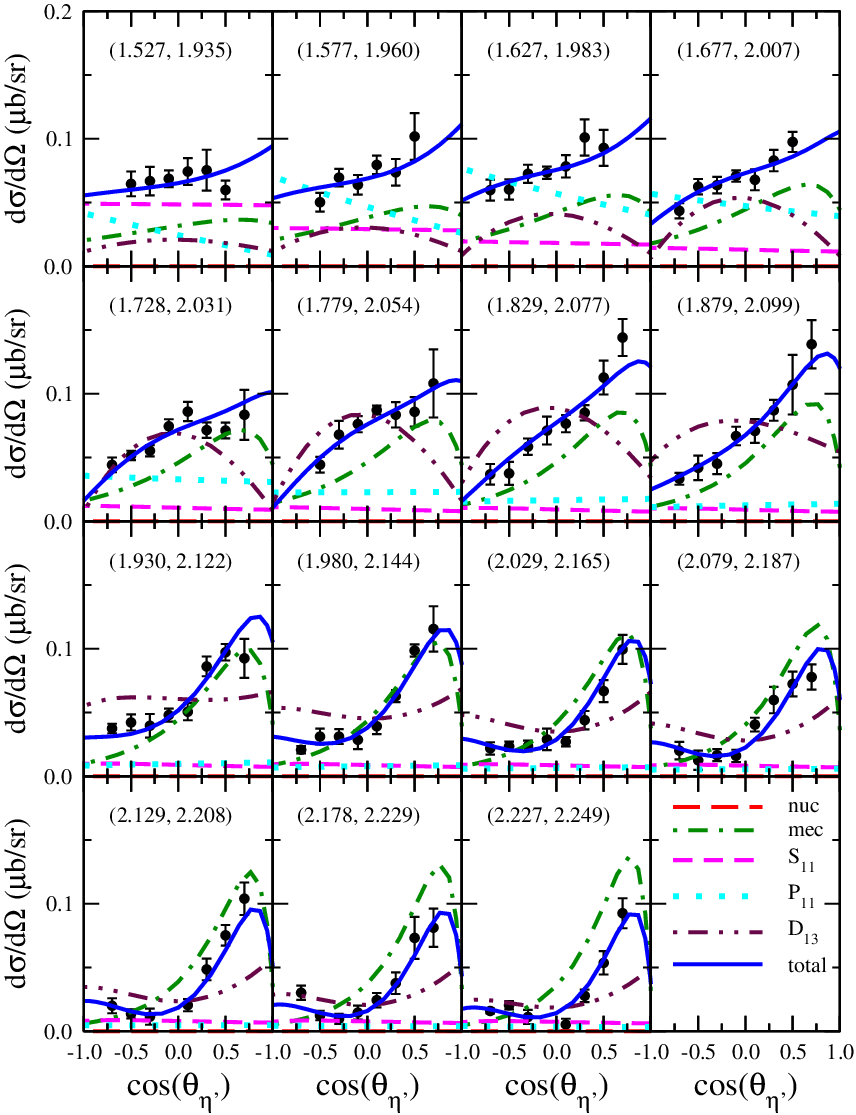}
\caption{\label{fig:R1428}%
(Color online) Same as in Fig.~\ref{fig:R1275} for the fit result of
Table~\ref{tbl:R1428}. The nucleonic current contribution (long-dashed curves)
is negligible and cannot be seen on the present scale.}
\end{figure}
%

\subsection{Differential cross sections}

\begin{figure}[t!]
\includegraphics[width=\columnwidth,angle=0,clip]{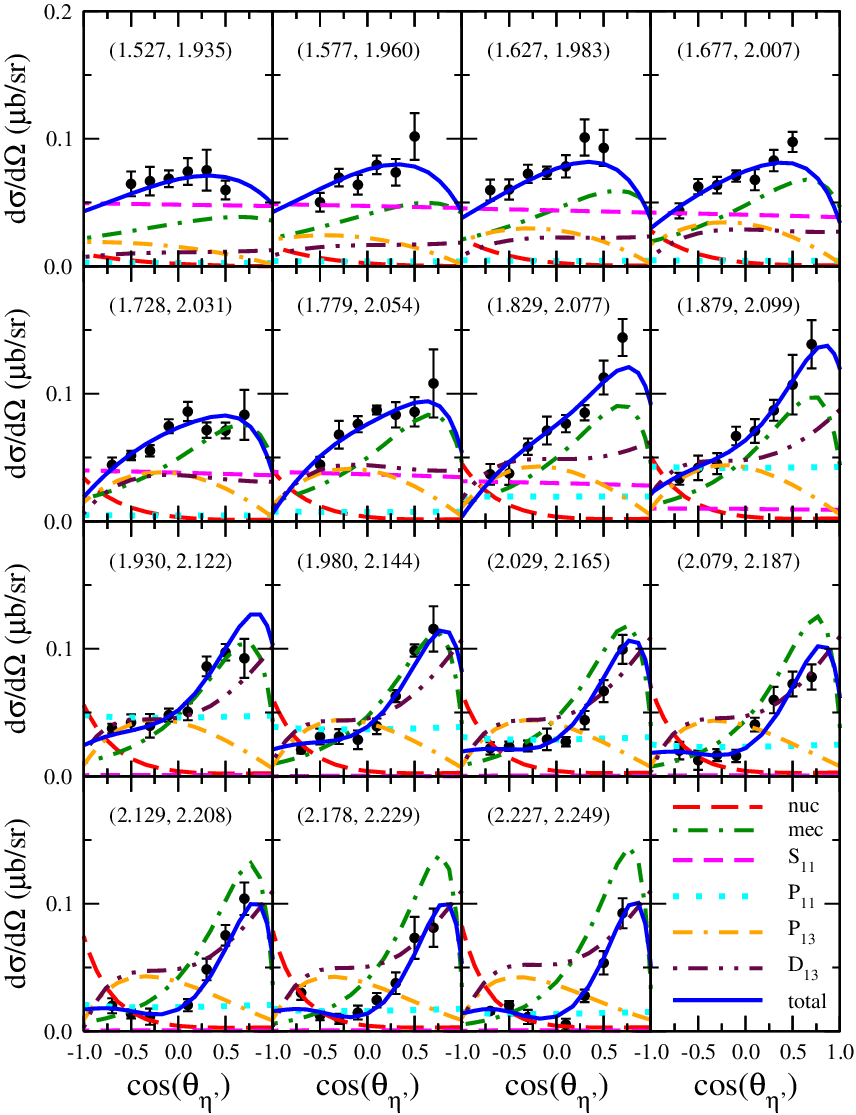}
\caption{\label{fig:R1447}%
(Color online) Same as in Fig.~\ref{fig:R1275} for the fit result of
Table~\ref{tbl:R1447}.}
\end{figure}

The details of the fits presented here are given in
Tables~\ref{tbl:R1275}--\ref{tbl:R1447} and the corresponding
Figs.~\ref{fig:R1275}--\ref{fig:R1447}. For the purpose of easy comparison,
Fig.~\ref{fig:Rall} provides an overview of all those results. All five fits
were obtained using the energy-dependent width functions of
Sec.~\ref{sec:widths}.
They all have comparable overall $\chi^2$ from each other and describe the data
quite well. We see that most of the differences among them are at forward and
backward angles where there are no data. Therefore, measurements of the cross
sections at more forward and backward angles than presently available would
disentangle some of the results in Fig.~\ref{fig:Rall}.
Despite the fact that the overall quality of the fits is comparable to each
other, the resulting parameter values are quite different. In particular, the
fit set in Table~\ref{tbl:R1275} contains the minimum number of resonances
(four) required to meet the present fit-quality criterium mentioned in the
beginning of this section. In contrast to the analysis of the SAPHIR data
\cite{NH1}, the inclusion of the spin-3/2 resonaces is important in order to
reproduce the data quantitatively. As mentioned before, although we cannot
identify the resonances uniquely in the present analysis, Table~\ref{tbl:R1275}
reveals that one of the resulting resonances, $P_{11}(2104)$, is consistent
with that quoted by the PDG~\cite{PDG} as one-star resonance.
In the fit set of Table~\ref{tbl:R1308}, we have included an additional $D_{13}$
resonance. Here, all the resonances but one are above the $\eta'$ production threshold
energy and that two of the resulting spin-3/2 resonances, $P_{13}(1900)$ and
$D_{13}(2084)$, are consistent with those seen and quoted by the PDG~\cite{PDG} as
two-star resonances. Also, in this particular set of parameters, the resulting
$NN\eta'$ coupling constant is very small.
The fit set of Table~\ref{tbl:R1515} includes three $S_{11}$ and three $P_{11}$
resonances, instead of one each as in Table~\ref{tbl:R1308}, keeping the number
of spin-3/2 resonances unchanged compared to the fit set of Table~\ref{tbl:R1308}.
Here, two of the $S_{11}$, one of the
$P_{11}$ and one of the $D_{13}$ resonances end up well below the production
threshold, while one $P_{11}$ resonance mass is close to 2.4 GeV. With the
exception of the latter resonance, all the resulting resonance masses are
consistent with those quoted by the PDG~\cite{PDG} as four-star
[$S_{11}(1535)$, $S_{11}(1650)$], three-star [$P_{11}(1710), D_{13}(1720)$],
two-star [$P_{13}(1900)$, $D_{13}(2080)$], and one-star [$S_{11}(2090),
P_{11}(2100)$] resonances. Here, the $NN\eta'$ coupling constant is
$g_{NN\eta'}\approx 1.3$.
In the fit result of Table~\ref{tbl:R1428}, we have omitted the $P_{13}$
resonance and considered two $S_{11}$, three $P_{11}$ and two $D_{13}$
resonances. Again, three of the resulting resonances, $S_{11}(1538)$,
$P_{11}(1710)$, and $D_{13}(2090)$, are consistent with known resonances.
The $NN\eta'$ coupling constant is practically zero, in line with the small
value obtained for the fit result of Table~\ref{tbl:R1308}.
 We have also considered all the
known spin-1/2 and \mbox{-3/2} resonances \cite{PDG} (including those with only
one star) in our fit.\footnote{There are
   also established spin-5/2, \mbox{-7/2}, and \mbox{-9/2}
   resonances \cite{PDG} in the energy region covered by the JLab data,
   but they have been omitted in the present analysis.}
The resulting parameter values are displayed
in Table~\ref{tbl:R1447}. Here the resonance masses are fixed at the respective
(centroid) values given in Ref.~\cite{PDG}. The resulting
resonance widths are all consistent with those quoted in Ref.~\cite{PDG}.
The $P_{11}(1440)$ resonance has practically no influence on the observables
considered here and, therefore, it has been omitted in the fit set shown.
For the $NN\eta'$ coupling constant, we obtained $g_{NN\eta'}=1.9$.

All these parameter sets illustrate the fact that cross sections do not impose
enough constraints to the fit in order to extract definitive information on the
resonances. Spin observables, on the other hand, do impose more stringent
constraints and help distinguish among these parameter sets, as we shall show
later.

\begin{table}[t!]
\begin{center}
\caption{Same as Table~\ref{tbl:R1275}. All resonance masses are kept at their
PDG values. $\chi^2/N=1.01$. }
\begin{tabular}{l@{\qquad}r@{\qquad}r@{\qquad}r@{\qquad}r}
\hline\hline
Nucleonic current: & & & & \\
$g_{NN\eta^\prime}$                      & 1.91           &                &                  &                  \\
$\lambda$                                & \textbf{0.0}   &                &                  &                  \\
$\Lambda_N$~(MeV)                        & \textbf{1200}  &                &                  &                  \\
\hline
Mesonic current: & & & &  \\
$g_{\eta^\prime\rho\gamma}$              & \textbf{1.25}  &                &                  &                  \\
$g_{\eta^\prime\omega\gamma}$            & \textbf{0.44}  &                &                  &                  \\
$\Lambda_v$ (MeV)                        &  1447          &                &                  &                  \\
\hline
$N^*=S_{11}$ current: & & & &  \\
$m_{N^*}$ (MeV)                          & \bf{1535}      &  \bf{1650}     & \bf{2090}        &                  \\
$g_{NN^*\gamma}\,g_{NN^*\eta^\prime}$    & -2.59          & 4.00           & -0.07            &                  \\
$\lambda$                                & 0.23           & 0.66           & 1.00             &                  \\
$\Lambda_{N^*}$ (MeV)                    & \textbf{1200}  &\textbf{1200}   &\textbf{1200}     &                  \\
$\Gamma_{N^*} $ (MeV)                    & 101            & 197            &   62             &                  \\
$\gamma_{N\gamma}  $                     &                &                & \textbf{0.001}   &                  \\
$\beta_{N\pi}  $                         & \textbf{0.5}   & \textbf{0.9}   &  0.04            &                  \\
$\beta_{N\eta}  $                        & \textbf{0.5}   & \textbf{0.1}   &                  &                  \\
$\beta_{N\eta^\prime} $                  &                &                &  0.96            &                  \\
\hline
$N^*=P_{11}$ current: & & & & \\
$m_{N^*}$ (MeV)                          &  \bf{1710}     &  \bf{2100}     &                  &                  \\
$g_{NN^*\gamma}\,g_{NN^*\eta^\prime}$    & -3.87          & -0.39          &                  &                  \\
$\lambda$                                & 0.27           & 0.14           &                  &                  \\
$\Lambda_{N^*}$ (MeV)                    & \textbf{1200}  & \textbf{1200}  &                  &                  \\
$\Gamma_{N^*} $ (MeV)                    &  249           &   75           &                  &                  \\
$\gamma_{N\gamma}  $                     &                &  \textbf{0.002}&                  &                  \\
$\beta_{N\pi}  $                         & \textbf{0.15}  & 0.50           &                  &                  \\
$\beta_{N\pi\pi}  $                      & \textbf{0.85}  &                &                  &                  \\
$\beta_{N\eta^\prime} $                  &                & 0.50           &                  &                  \\
\hline
$N^*=P_{13}$ current: & & & &  \\
$m_{N^*}$ (MeV)                          &  \bf{1720}     &  \bf{1900}     &                  &                  \\
$g_{1NN^*\gamma}\,g_{NN^*\eta^\prime}$   &  -0.44         &  0.04          &                  &                  \\
$g_{2NN^*\gamma}\,g_{NN^*\eta^\prime}$   & 1.49           & -0.54          &                  &                  \\
$\Lambda_{N^*}$ (MeV)                    & \textbf{1200}  & \textbf{1200}  &                  &                  \\
$\Gamma_{N^*} $ (MeV)                    &  107           &  316           &                  &                  \\
$\beta_{N\pi}  $                         & \textbf{0.2}   & \textbf{0.6}   &                  &                  \\
$\beta_{N\rho}  $                        & \textbf{0.8}   &                &                  &                  \\
$\beta_{N\omega}  $                      &                & \textbf{0.4}   &                  &                  \\
\hline
$N^*=D_{13}$ current: & & & & \\
$m_{N^*}$ (MeV)                          & \textbf{1520}  & \textbf{1700}  & \textbf{2080}    &                  \\
$g_{1NN^*\gamma}\,g_{NN^*\eta^\prime}$   & -1.00          & -0.92          & -0.07            &                  \\
$g_{2NN^*\gamma}\,g_{NN^*\eta^\prime}$   & 0.46           & 0.95           & 0.08             &                  \\
$\Lambda_{N^*}$ (MeV)                    & \textbf{1200}  & \textbf{1200}  & \textbf{1200}    &                  \\
$\Gamma_{N^*} $ (MeV)                    & 135            &   49           & 102              &                  \\
$\gamma_{N\gamma}  $                     &                &                &\textbf{0.001}    &                  \\
$\beta_{N\pi}  $                         & \textbf{0.55}  & \textbf{0.10}  & 0.87             &                  \\
$\beta_{N\pi\pi}  $                      & \textbf{0.45}  & \textbf{0.90}  &                  &                  \\
$\beta_{N\eta^\prime} $                  &                &                & 0.13             &                  \\
\hline\hline
\end{tabular}
\label{tbl:R1447}
\end{center}
\vspace{-2ex}
\end{table}
%
%

Although the parameter sets in Tables~\ref{tbl:R1275}--\ref{tbl:R1447} yield
comparable fits to the cross section, the corresponding dynamical contents are
quite different from each other. Let us discuss, therefore, some of the
different features present in the results corresponding to the various parameter
sets. Figure~\ref{fig:R1275} shows some details of the dynamical content of our
model corresponding to the fit results given in Table~\ref{tbl:R1275}. Here, both
the $S_{11}$ and $P_{13}$ resonances have the largest contribution at low energies;
the former dies out as the energy increases while the latter contribution persists
to higher energies. The angular shape of the $P_{13}$ resonance current
contribution is concave with a maximum at $\theta_{\eta'} \approx 90^\circ$. The
$P_{11}$ resonance contributes mostly around $T_\gamma=1.879$ GeV. Its angular
shape is rather flat (note that it includes both the $s$- and $u$-channel
contributions). However, its interference with other contributions, such as that due
to the $S_{11}$ resonance, leads to a distinctive angular dependence. Although the
$D_{13}$ resonance current is relatively small in this particular fit set, it plays
an important role in reproducing the data through its interference with other currents.
The mesonic current contribution plays a crucial role in reproducing the observed
forward-peaked angular distribution, especially at higher energies. This is a general
feature observed in many reactions at high energies where the $t$-channel mechanism
(either Regge trajectories or meson exchanges) accounts for the small-$t$ behavior of
the cross section. However, the present result shows also a competing mechanism due
to resonances and that the observed forward-peaked angular distribution is a result
of significant interference effects. We note that this feature is not restricted to
the particular set of the parameter values of Table~\ref{tbl:R1275}, but it is also
found in other sets that fit the data (note, in particular, a relatively large
$S_{11}$ resonance contribution in Fig.~\ref{fig:R1515} and a $D_{13}$ resonance
contribution in Figs.~\ref{fig:R1308}, \ref{fig:R1428} and \ref{fig:R1447} at higher
energies). Therefore, this feature prevents us from fixing uniquely the mesonic
current from the cross-section data at forward angles and higher energies.
The nucleonic current contribution is very small here; however, as mentioned in
the beginning of this section, the cross-section data alone do not impose stringent
constraints on the fit so that it is possible to reproduce the data equally well with
a much larger coupling constant, as can be seen in Tables~\ref{tbl:R1515} and
\ref{tbl:R1447}. We will come back to this issue later, in
Sec.~\ref{sec:etaprimecoupling}.

Figure~\ref{fig:R1308} shows the dynamical content of our model corresponding
to the fit results given in Table~\ref{tbl:R1308}. Here, the $S_{11}$ resonance
contribution is considerable only at the lowest energy measurement and already
at $T_\gamma=1.627$\,GeV, it becomes very small and it is practically
negligible for higher energies. Both the $P_{11}$ and $P_{13}$ resonance
contributions exhibit similar features to those observed in
Fig.~\ref{fig:R1275}, with both contributions reaching its maximum around
$T_\gamma=1.677$\,GeV. The $D_{13}$ resonance contribution is large over the
energy region considered, except at lower energies. Above $T_\gamma\approx
1.677$\,GeV, it is the largest contribution. Its angular shape changes
drastically with energy, starting with a small negative curvature in the lower
energy region and ending with a roughly convex shape in the higher energy
region. Note that this energy dependence of the angular shape is due to an
interference between the two $D_{13}$ resonances with different masses.
Although somewhat larger, the mesonic current contribution is essentially the same
to that in Fig.~\ref{fig:R1275}. As has been pointed out above, there is a
considerable interference effects between the mesonic and the ($D_{13}$) resonance
currents at higher energies. The nucleonic current is practically zero in
Fig.~\ref{fig:R1308} since, as mentioned above, the resulting $NN\eta'$ coupling
constant is very small.

In the fit result of Table~\ref{tbl:R1515} shown in Fig.~\ref{fig:R1515}, the
$S_{11}$ resonance contribution is very strong especially in the lower energy
region and is quite appreciable even at higher energies. The $P_{11}$ resonance
contribution basically shows the same feature as in the fit sets discussed above.
The $P_{13}$ resonance contribution exhibits a convex angular shape, just
opposite to the concave shape shown in Figs.~\ref{fig:R1275} and
\ref{fig:R1308}. This difference is due to the relative sign difference
between the coupling constants $g_{1NN^*\eta'}$ and $g_{2NN^*\eta'}$ as
compared to the results shown in Figs.~\ref{fig:R1275} and \ref{fig:R1308}.
The $D_{13}$ resonance contribution is largest around
$T_\gamma=1.829$\,GeV. The shape of the angular distribution is quite different
from the other fit discussed above. Together with the $P_{13}$ resonance, it
describes some of the details of the observed angular distribution around
$T_\gamma=1.779$\,GeV. The mesonic current is much larger in this fit than in
the other fits. In particular, it largely overestimates the measured cross
section in the higher energy region. Its destructive interference with the
other currents brings down the total contribution in agreement with the data.
Once more, this shows that one has to be cautious in trying to fix the
$t$-channel contribution using the cross-section data at forward angles and
higher energies. The nucleonic current gives an appreciable contribution in
this fit, especially at higher energies and backward angles due to the
$u$-channel. The corresponding $NN\eta'$ coupling constant is
$g_{NN\eta'}=1.33$.

Figure~\ref{fig:R1428} shows the dynamical content of the fit result of
Table~\ref{tbl:R1428}. The $S_{11}$ resonance contribution is largest at the
lowest energy, but it decreases quickly as the energy increases. The $P_{11}$
resonance contribution is largest around $T_\gamma=1.627$\,GeV, with more
pronounced angular distribution than in the other fit results. The $D_{13}$
resonance contribution has a concave angular shape in the lower energy region
and is largest at around $T_\gamma=1.829$\,GeV. For higher energies the angular
shape changes and gives the largest contribution for forward angles apart from
the mesonic current, the latter providing again the bulk of the observed rise
of the cross section at forward angles. The $P_{13}$ resonance is not included
in this fit set. We note that, unlike in Fig.~\ref{fig:R1515}, some of the
details of the observed angular distribution around $T_\gamma=1.779$\,GeV is
not well reproduced, indicating the importance of both the $P_{13}$ and
$D_{13}$ resonances. The nucleonic current contribution is practically zero.

The dynamical content of the fit result given in Table~\ref{tbl:R1447} is shown
in Fig.~\ref{fig:R1447}. Overall, except for the lower energies, the mesonic
current yields the largest contribution. The $S_{11}$ and $P_{13}$ resonance
contributions are important in the low energy region while the $P_{11}$,
$P_{13}$, and $D_{13}$ resonances are important in the higher energy region.
The nucleonic current is non-negligible only for backward angles at higher
energies. Here, one major difference from the other fit results is the rather
pronounced bending downward of the cross section (solid curves) for forward
angles at lower energies. Measurements of the cross sections for more forward
angles would tell us whether such a behavior would indeed be necessary.

\begin{figure}[t!]\centering
\includegraphics[width=.65\columnwidth,angle=0,clip]{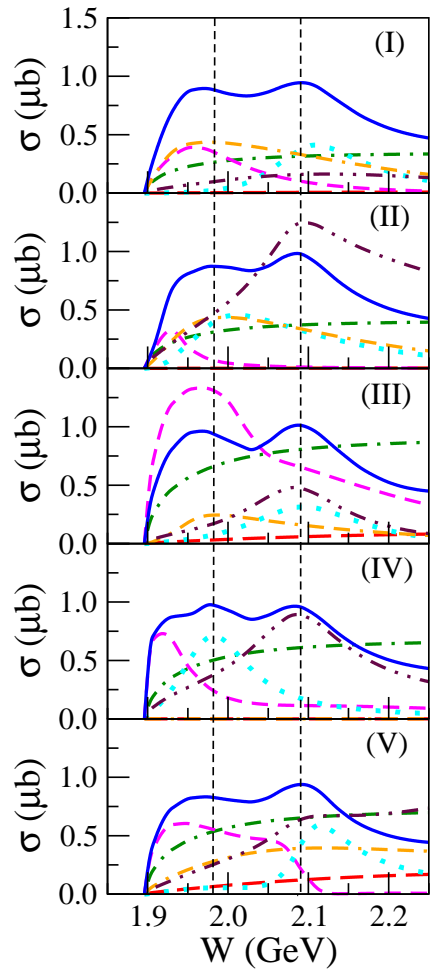}
\caption{\label{fig:Ralltot}%
(Color online) Total cross section for $\gamma p\to p \eta'$ as a function of
the total energy of the system, $W=\sqrt{s}$. The panels from top to bottom
correspond to the fit results of Tables~\ref{tbl:R1275}--\ref{tbl:R1447}, as
indicated.
The overall total cross sections
(solid lines) are broken down according to their dynamical contributions, with
line styles defined in Fig.~\ref{fig:R1275}. The two dashed vertical lines are
placed to guide the eye through the two bump positions in all panels.}
\end{figure}
%

\subsection{Total cross sections}

Figure~\ref{fig:Ralltot} shows the predictions for the total cross sections
obtained by integrating the corresponding differential cross sections shown in
Figs.~\ref{fig:R1275}-\ref{fig:R1447}. Although these predictions may suffer
from considerable uncertainties due to differences in the corresponding
differential cross sections, especially at lower energies, they exhibit a common
feature, i.e., the total cross sections (solid curves) seems to show a bump
structure around $W=2.09$\,GeV which is caused mainly by the $P_{11}$ and/or
$D_{13}$ resonance depending on the fit set.
Note that the PDG quotes a two-star $D_{13}(2080)$ and a one-star
$P_{11}(2100)$ resonance which are practically at this bump position. There is
also a one-star resonance, $S_{11}(2090)$, that is just at the bump and,
therefore, might have contributed to its structure. However, the angular
distribution does not favor this possibility. The total cross section also
seems to exhibit a bump structure at a lower energy of around $W=1.96$\,GeV due
to the $S_{11}$, $P_{11}$ and/or $P_{13}$ resonance depending on the fit set.
The latter two resonances can also contribute to the broadening of this bump
depending on the fit set, as can be seen in Fig.~\ref{fig:Ralltot}. A rather
sharp rise of the cross section from the threshold is caused by the $S_{11}$
resonance, except in the top two panels, where the $P_{13}$ resonance also
contributes to this rise. The structures exhibited by the total cross section,
in particular the bump around $W=2090$ GeV, are unlikely to be artifacts of the
present predictions and, consequently, we would expect them to show up in the
actual total cross-section data when they are measured.

\subsection{\boldmath $NN\eta'$ coupling constant}\label{sec:etaprimecoupling}

\begin{figure}[b!]\centering
\includegraphics[width=.85\columnwidth,angle=0,clip]{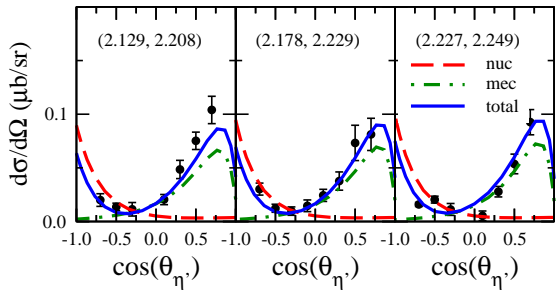}
\caption{\label{fig:NoRes}%
(Color online) Fit result with no resonances. $g_{NN\eta^\prime}=2.10$ and
$\Lambda_v=1264$\,MeV. See caption of Fig.~\ref{fig:R1308}.} \vspace{-10bp}
\end{figure}

As we have seen in this section, unfortunately the present analysis cannot
determine the $NN\eta'$ coupling constant, since the available cross-section
data can be reproduced equally well with different sets of parameters in which
this coupling constant varies considerably. However, an upper limit of its
value can still be estimated. One of the reasons why $g_{NN\eta'}$ cannot be
extracted uniquely from the cross-section data is that the resonance currents,
especially the one due to the $D_{13}$ resonance, can give rise to the observed
enhancement of the backward-angle cross section as shown in
Figs.~\ref{fig:R1308} and \ref{fig:R1428}. Also, the $P_{11}$ resonance current
alone can lead to a feature of the cross section similar to that due to the
nucleonic current, i.e., the enhancement of the backward-angle cross section at
higher energies through the $u$-channel contribution. The resonance currents
can also interfere destructively with the nucleonic current in which case one
obtains a larger $NN\eta'$ coupling constant. In fact, in a very extreme case,
we have obtained a fit value as large as $g_{NN\eta'}=3.0$. It is obvious,
therefore, that a more unambiguous extraction of this coupling constant
requires going to an energy region where the resonance contributions are small.
Figure~\ref{fig:NoRes} illustrates this point; here we show the fit result
considering the data with energies at $T_\gamma=2.129$\,GeV and above only and
assuming a scenario in which no resonance currents contribute at these
energies. The resulting fit parameters are $g_{NN\eta^\prime}=2.10$ for the
$NN\eta^\prime$ coupling constant and $\Lambda_v=1264$\,MeV for the cutoff
parameter in the form factor at the $\eta' v\gamma$ vertex. In
Fig.~\ref{fig:NoRes} we see that the nucleonic and mesonic currents interfere
with each other. However, the interference pattern is such that it does not
cause any problem in fixing both the nucleonic and mesonic current
contributions to a large extent. In any case, judging from the overall results
of our analysis, we would not expect $g_{NN\eta'}$ to be much larger than 2.

%

\subsection{Meson exchanges versus Regge trajectory}

It is well known that $t$-channel processes at high energies (above
$\sim$3--5\,GeV) may also be described by Regge trajectories. However, how far
down in energy one can go with this description before an explicit inclusion of
ordinary meson exchange is required is still an open issue. We would expect a
smooth transition from a description in terms of meson exchanges to one in
terms of Regge trajectory as one goes up higher in energy. The issue of meson
exchanges versus Regge trajectory is particularly relevant in the present
context, for the extracted resonance parameters can depend on these two
alternatives for modeling the $t$-channel contribution \cite{NH1}.

In their analysis of the SAPHIR data \cite{SAPHIR}, Chiang \emph{et
al.}~\cite{Chiang} advocate the use of Regge trajectories while other authors
\cite{others} have employed vector-meson exchanges. In our previous analysis of
these data, we found \cite{NH1} that they can be reproduced equally well using
either meson exchanges (with form factors at the $v\eta'\gamma$ vertices) or a
Regge trajectory for the $t$-channel contribution. It is interesting to see
whether the same is true for the new CLAS data \cite{CLAS}. We have repeated
the calculation with the Regge trajectory following Ref.~\cite{NH1}. In
particular, we replace the $t$-channel meson exchange propagators by the
corresponding Regge trajectories keeping everything else unchanged, except that
the form factor at the $v\eta'\gamma$ vertex is set to unity. All the free
parameters of the model are refitted again. We found that the fit quality using
the Regge trajectory is, at best, comparable to that obtained using the
ordinary meson exchanges for the $t$-channel. For example, the $\chi^2$
corresponding to the fit set of Table~\ref{tbl:R1308} is $\chi^2/N=1.17$
compared to $\chi^2/N=1.04$ obtained with explicit meson exchanges; for the fit
set of Table~\ref{tbl:R1447}, the Regge-trajectory result is $\chi^2/N=1.08$,
as compared to $\chi^2/N=1.01$ using meson exchanges. Similar results are
obtained for other fit sets considered in subsection III.A.

\subsection{Spin observables}

\begin{figure}[t!]
\includegraphics[width=\columnwidth,angle=0,clip]{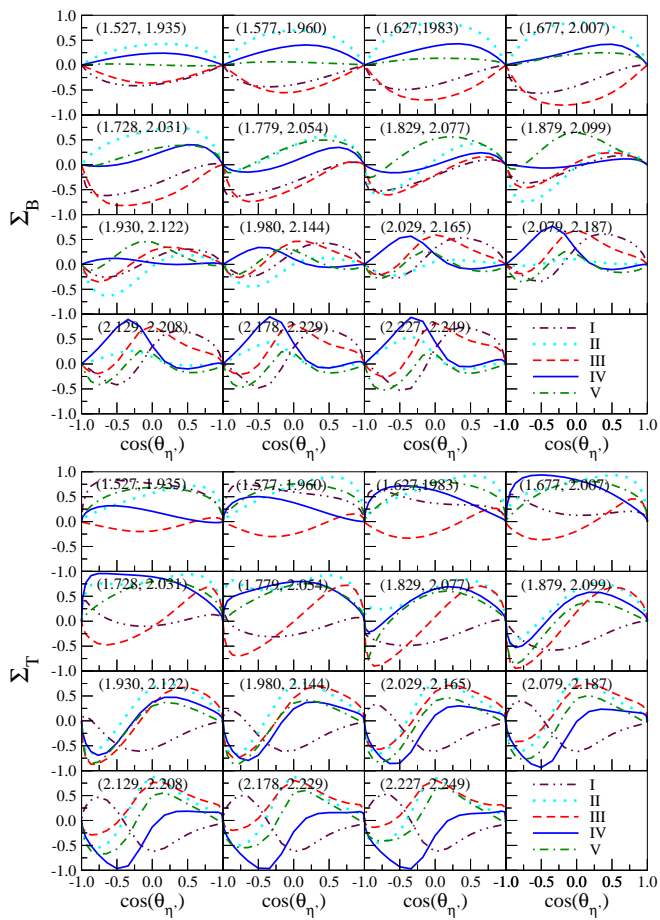}
\caption{\label{fig:spinBT}%
(Color online) Photon beam and target nucleon asymmetries $\Sigma_B$ (top
panel) and $\Sigma_T$ (bottom), respectively, for $\gamma p\to p \eta'$ as a
function of the $\eta'$ emission angle $\theta_{\eta'}$ in the
center-of-momentum system. See the caption of Fig.~\ref{fig:Rall} for the
meaning of the different curves. }
\end{figure}

We now turn our attention to spin observables. As we have shown in
Fig.~\ref{fig:Rall}, cross sections do not impose very severe constraints on
the model parameter values. We expect spin observables to be more sensitive in
this respect. The predictions for the beam and target asymmetries corresponding
to the fit results of Tables~\ref{tbl:R1275}--\ref{tbl:R1447} are shown in
Fig.~\ref{fig:spinBT}. As we can see, unlike the cross sections (see
Fig.~\ref{fig:Rall}), the predictions vary considerably between the different
parameter sets. For energies where the beam asymmetry is less sensitive to the
parameter sets, the target asymmetry is quite sensitive and \emph{vice versa}.
Therefore, overall, a combined analysis of these spin observables will impose
much more stringent constraints on the fit and should help determine better the
model parameters.

%

\section{Summary}

We have analyzed the new CLAS~\cite{CLAS} data of the $\gamma p \to p
\eta^\prime$ reaction within an approach based on a relativistic meson-exchange
model of hadronic interactions. The present model is an extension of the one
reported in Ref.~\cite{NH1} and it includes the nucleonic and the mesonic as
well as the nucleon-resonance currents. The latter includes both spin-1/2 and
\mbox{-3/2} resonance contributions in contrast to our previous
work~\cite{NH1}, where only spin-1/2 resonances were considered. In addition,
we employ energy-dependent resonance widths in the present work. The resulting
reaction amplitude is fully gauge invariant.

We have shown that the mesonic as well as the spin-1/2 and \mbox{-3/2}
resonance currents are important to describe the existing data quantitatively.
The observed angular distribution is due to delicate interference effects
between the different currents. In our analysis, most of the resulting
resonances may be identified with known resonances~\cite{PDG}. We emphasize,
however, that one should be cautious with such an identification of the
resonances. As we have seen, the cross-section data alone do not impose enough
constraints for an unambiguous determination of the resonance parameters. In
this connection, we have shown that the beam and target asymmetries can help
impose more stringent constraints. Furthermore, there is a possibility that
some of the resonances in the present work are mocking up background
contributions, especially those due to the final-state interaction, which is
not taken into account explicitly in our calculation. Obviously, effects of the
final-state interaction should be investigated in future work before a
conclusive identification of the resonances can be made.

We have predicted a bump structure in the total cross section at $W\approx
2.09$ GeV (see Fig.~\ref{fig:Ralltot}). If this is confirmed, the
$D_{13}(2080)$ and/or $P_{11}(2100)$ resonance may be responsible for this
bump.

Our study also shows that the nucleonic current should be relatively small.
However, contrary to the expectation in our earlier work~\cite{NH1}, the new
high-precision cross-section data do not allow to pin down this current
contribution due to the possible presence of resonance currents, especially of
the $D_{13}$ resonance, which can also lead to an enhancement of the cross
section for backward angles at higher energies, a feature that otherwise arises
from the $u$-channel nucleonic current contribution. These complications
notwithstanding, assuming that for the very high end of the present data set
resonance contributions can be neglected, we argue in
Sec.~\ref{sec:etaprimecoupling} that the upper limit of $g_{NN\eta^\prime}$ can
now be lowered to a value of $g_{NN\eta^\prime}\lesssim 2$, whereas our
previous analysis \cite{NH1} had suggested an upper limit of
$g_{NN\eta^\prime}\lesssim 3$. Further corroboration of this finding is needed.

In this respect, it should be noted that the result pertaining here to the
$NN\eta^\prime$ coupling constant is, of course, a model-dependent one. Indeed,
what is relevant in our calculations is the product of $g_{NN\eta^\prime}$ and
the associated hadronic form factor which accounts for the off-shellness of the
intermediate nucleon. Moreover, our $NN\eta^\prime$ coupling constant is
defined at the on-mass-shell point, i.e.,
$g_{NN\eta^\prime}=g_{NN\eta^\prime}(q^2=m_{\eta'}^2)$ while the coupling
required in Eq.~(\ref{spinfrac}) in connection with the origin of the nucleon
spin is at $q^2=0$. Since the $\eta'$ meson is a relatively heavy meson
($m_{\eta'}\approx 957$\,MeV), we would expect that $g_{NN\eta^\prime}(q^2=0)$
will be considerably smaller than its value at $q^2=m_{\eta'}^2$ because of the
presence of the form factor which usually cut down the coupling strength.
Therefore, we might well expect that the $NN\eta'$ coupling at $q^2=0$ to be
negligibly small, consistent with zero.

We have also shown that the mesonic current contribution cannot be fixed
unambiguously from the existing cross-section data because of the possible
presence of the resonance currents, especially the $D_{13}$ resonance. A
possibility to determine the $t$-channel current is to measure the cross
sections at higher energies where the resonance contributions becomes
negligible.

Furthermore, we have found that using a Regge trajectory in the $t$-channel
instead of explicit meson exchanges yields overall fit qualities that are, at
best, comparable to those obtained with meson exchanges. This indicates that
explicit $\rho$ and $\omega$ exchanges, as employed here, are completely
adequate to describe the $t$-channel degrees of freedom at the present
energies.

Finally, the results of the present work should provide useful information for
further investigations, both experimentally and theoretically, of the $\gamma N
\to N \eta^\prime$ reaction. In particular, measurements of cross sections at
smaller forward and larger backward angles than are available in the present
data set would already help constrain the model parameters considerably, as can
be seen in Fig.~\ref{fig:Rall}. Total cross sections should also be measured in
order to confirm or dismiss the bump structures, especially around $W=2.09$
GeV, predicted in the present calculation. In addition, it is expected that
measurements of spin observables --- such as beam and target asymmetries shown
in Fig.~\ref{fig:spinBT} --- would impose more stringent limits on the range of
permissible parameters and this would undoubtedly provide a much improved
description of the resonances and their properties in the energy region covered
by the existing data. From the theoretical side, it is possible that the
nucleon resonances introduced in the present work are mocking up the background
contributions not taken into account in the calculation. In this connection, it
is extremely interesting to investigate effects of the final-state interaction
which has not been treated explicitly in the present calculation.
Unfortunately, at present no realistic model is available that can provide the
relevant $\eta' N$ final-state interaction. In addition, effects of higher-spin
resonances that have been ignored in the present analysis should be
investigated in the future.

\begin{acknowledgments}
The authors thank M. Dugger, B.\,G.~Ritchie, and the CLAS Collaboration for
providing the $\eta'$ data prior to publication. This work was supported by the
COSY Grant No.\ 41445282\;(COSY-58).
\end{acknowledgments}

\appendix

\section{Spin-3/2 Resonance Propagator}

We employ here the Rarita--Schwinger (RS) choice for the free Lagrangian of a
spin-3/2 particle with mass $m$,
\begin{equation}
\mathcal{L} = \bar{\psi}_\mu \Lambda^{\mu\nu} \psi_\nu~,
\end{equation}
where
\begin{equation}
\Lambda^{\mu\nu}=-\frac{i}{2}\left[\sigma^{\mu\nu},(\fs{p}-m)\right]_+~,
\end{equation}
with $p=i\partial$, the anticommutator bracket $[a,b]_+=ab+ba$, and
$\sigma^{\mu\nu}=\frac{i}{2}(\gamma^\mu\gamma^\nu-\gamma^\nu\gamma^\mu)$. (In
the RS choice, the parameter $A$ that usually appears in $\mathcal{L}$ is taken
as $A=-1$ \cite{RSprop}.) From
\begin{equation}
\Lambda_{\mu\lambda}S^{\lambda\nu}=S_{\mu\lambda}\Lambda^{\lambda\nu}
=g_\mu^{\;\nu}~,
 \label{eq:propcond}
\end{equation}
the propagator is then found as
\begin{equation}
S^{\mu\nu}(p)
 = \frac{(\fs{p}+m)\Delta^{\mu\nu}}{p^2-m^2}
 =\frac{\tilde{\Delta}^{\mu\nu}(\fs{p}+m)}{p^2-m^2}~,
 \label{eq:S3prop}
\end{equation}
where $\Delta^{\mu\nu}$ is the RS tensor of Eq.~(\ref{eq:RStensor}) (with
$m=m_R$) and
\begin{equation}
\tilde{\Delta}^{\mu\nu}
 =-g^{\mu\nu}+\frac{1}{3}\gamma^\mu\gamma^\nu
        + \frac{2p^\mu p^\nu}{3m^2}
                 -\frac{\gamma^\mu p^\nu-p^\mu\gamma^\nu}{3m}~,
\label{eq:RStensor1}
\end{equation}
which differs from (\ref{eq:RStensor}) by the sign of the last term.

When seeking an ansatz for describing a spin-3/2 resonance, we note first that
there are, of course, infinitely many ways to achieve a pole description whose
on-shell behavior \emph{on the real axis} corresponds to replacing the mass of
the elementary propagator by
\begin{equation}
m\to m_R -i\frac{\Gamma_R}{2}~,
\end{equation}
where $m_R$ is the resonance mass and $\Gamma_R$ the associated width. In
constructing a resonant propagator, we are guided by the following motivation.
As in the spin-1/2 case of (\ref{eq:spin1}), we want to describe the spin in
terms of the elementary operators, i.e., we want to preserve the numerator
structure of (\ref{eq:S3prop}) \emph{and} the symmetry between the RS tensors
$\Delta$ and $\tilde{\Delta}$. In a schematic matrix notation, we therefore
make the ansatz
\begin{equation}
S=X^{-1} (\fs{p}+m)\Delta = \tilde{\Delta}(\fs{p}+m)\tilde{X}^{-1}~,
 \label{eq:Deltacondition}
\end{equation}
putting, in analogy to the denominator of the spin-1/2 case (\ref{eq:spin1}),
\begin{subequations}
\begin{align}
X&=(p^2-m_R^2)g +i A(\fs{p}+m_R)~,
\\[1ex]
\tilde{X}&=(p^2-m_R^2)g+i(\fs{p}+m_R)\tilde{A}~,
\end{align}
\end{subequations}
with the operators $A$ and $\tilde{A}$ to be determined such that the second
equality in (\ref{eq:Deltacondition}) holds true, i.e.,
\begin{equation}
X^{-1} (\fs{p}+m)\Delta = \tilde{\Delta}(\fs{p}+m)\tilde{X}^{-1}~.
\end{equation}
Multiplying this equation by $\Lambda$ from both sides, one immediately finds
the condition
\begin{equation}
(\fs{p}+m_R)\tilde{A}\Lambda = \Lambda A (\fs{p}+m_R)~.
\end{equation}
In view of Eq.~(\ref{eq:propcond}) and the fact that on-shell, at $p^2=m_R^2$
and acting on a spin-3/2 eigenstate, the propagator must provide the width
information, we find that the ansatz
\begin{equation}
\tilde{A}^{\mu\nu}=-\Delta^{\mu\nu}\frac{\Gamma}{2}
 \quad\text{and}\quad
 A^{\mu\nu}= -\tilde{\Delta}^{\mu\nu}\frac{\Gamma}{2}~,
\end{equation}
satisfies all constraints. $\Gamma$ here may be any conveniently chosen width
function that goes to the static width $\Gamma_R$ at the resonance mass $m_R$.
We thus have
\begin{subequations}
\begin{equation}
S(p)=\left[(\fs{p}-m_R)g-i\frac{\Delta}{2}\Gamma\right]^{-1}\Delta
\end{equation}
or
\begin{equation}
S(p)=\tilde{\Delta}\left[(\fs{p}-m_R)g-i\frac{\tilde{\Delta}}{2}\Gamma\right]^{-1}~.
\end{equation}
\end{subequations}
By construction, both forms are completely equivalent, similar to the
equivalence of both forms for the elementary propagator (\ref{eq:S3prop}).

The inversion here is to be performed on the full 16-dimensional space of
Lorentz indices and component indices. There are various equivalent ways to do
this; we have done it by introducing indices
\begin{equation}
   i = 4\mu+\beta  \quad\text{and}\quad  j = 4\nu+\alpha~,
\end{equation}
where $\mu,\nu=0,1,2,3$ are the Lorentz indices and $\beta,\alpha=1,2,3,4$ are
the component indices, and defining $16\times 16$ numerator and denominator
matrices by
\begin{align}
N_{ij} &= \Delta^{\mu\lambda}_{\beta\alpha}g_{\lambda\nu}
 \nonumber\\
 &=-\delta_{\mu\nu}\delta_{\beta\alpha}
 + \frac{2p^\mu p_\nu}{3m_R^2}\delta_{\beta\alpha}
  \nonumber\\
 &\qquad\mbox{}
 +\left(\frac{1}{3}\gamma^\mu_{\beta\varepsilon}\gamma^\lambda_{\varepsilon\alpha}
                 +\frac{\gamma^\mu_{\beta\alpha}
                 p^\lambda-p^\mu\gamma^\lambda_{\beta\alpha}}{3m_R}\right)g_{\lambda\nu}
\end{align}
and
\begin{align} D_{ij}=p_\lambda
\gamma^\lambda_{\beta\alpha}\delta_{\mu\nu}
 -m_R\delta_{\mu\nu}\delta_{\beta\alpha}-iN_{ij}\frac{\Gamma}{2}~,
\end{align}
respectively. Numerically inverting the denominator matrix $D$, we then
calculate the spin-3/2 propagator as
\begin{equation}
S^{\mu\nu}_{\beta\alpha} = \left(D^{-1} N \right)_{ik}g^{\rho\nu}~,
\end{equation}
where $k=4\rho+\alpha$ and summation over $\rho$ is implied, as usual.



\begin{thebibliography}{99}
\bibitem{Capstick1}
S. Capstick and N. Isgur, Phys.\ Rev.\  D\,\textbf{34}, 2809 (1986); S.
Capstick and W. Roberts, \textit{ibid.} \textbf{47}, 1994 (1993);
                                           \textbf{49}, 4570 (1994);
                                           \textbf{57}, 4301 (1998);
                                           \textbf{58}, 074011 (1998).

\bibitem{EMC88}
J. Ashman \textit{et al.}, Phys.\ Lett.\ \textbf{B206}, 364 (1988).

\bibitem{Shore}
G.\,M. Shore and G. Veneziano, Nucl.\ Phys.\ \textbf{B381}, 23 (1992).

\bibitem{Efremov}
T. Hatsuda, Nucl.\ Phys.\ \textbf{B329}, 376 (1990); A.\,V. Efremov, J. Soffer,
and N. A. T\"ornqvist, Phys.\ Rev.\   Lett.\ \textbf{64}, 1495 (1990); Phys.\
Rev.\  D~\textbf{44}, 1369 (1991).

\bibitem{Venez1}
G.\,M. Shore and G. Veneziano, Phys.\ Lett.\ \textbf{B244}, 75 (1990).

\bibitem{Feldmann}
T. Feldmann, Int.\ J.\ Mod. \ Phys. \ \textbf{A15}, 159 (2000).


\bibitem{OZI}
S. Okubo, Phys.\ Lett.\ \textbf{5}, 165 (1963); G. Zweig, CERN Report No.\
TH412, 1964; J. Iizuka, Prog.\ Theor.\ Phys.\ Suppl.\ \textbf{37}--\textbf{38},
21 (1966).

\bibitem{SMC97}
D. Adams \textit{et al.}, Phys.\ Rev. \ \textbf{D56}, 5330 (1997).

\bibitem{x1}
G. Altarelli and G.\,G. Ross, Phys.\ Lett.\ \textbf{B212}, 391 (1988); R.\,D.
Carlitz, J.\,C. Collins, and A.\,H. M\"uller, \textit{ibid.} \textbf{B214}, 229
(1988).

\bibitem{NH1}
K.~Nakayama and H.~Haberzettl, Phys.\ Rev.\ C\,\textbf{69}, 065212 (2004).

\bibitem{CLAS}
M. Dugger \textit{et al.} (CLAS collaboration), nucl-ex/0512019.

\bibitem{SAPHIR}
R. Pl\"otzke \textit{et al.}, Phys.\ Lett.\ \textbf{B444}, 555 (1998); J. Barth
\textit{et al.}, Nucl.\ Phys.\ \textbf{A691}, 374c (2001).

\bibitem{hh97g}
H.~Haberzettl, Phys.\ Rev.\  C~\textbf{56}, 2041 (1997).

\bibitem{hhtree98}
H.~Haberzettl, C.~Bennhold, T.~Mart, and T.~Feuster, Phys.\ Rev.\
C~\textbf{58}, R40 (1998).

\bibitem{dw2}
R.\,M.~Davidson and R.~Workman, Phys.\ Rev.\  C~\textbf{63}, 025210 (2001).

\bibitem{walker}
R.\,L.~Walker, Phys.\ Rev.\ \textbf{182}, 1729 (1969).

\bibitem{arndt90}
R.\,A.~Arndt, R.\,L. Workman, Z.~Li, and L.\,D.~Roper, Phys.\ Rev.\
C~\textbf{42}, 1864 (1990).

\bibitem{lvov97}
A.\,I.~L'vov, V.\,A.~Petrun'kin, and M.~Schumacher,
 Phys.\ Rev.\ C\,\textbf{55}, 359 (1997).

\bibitem{drechsel}
D.~Drechsel, O.~Hanstein, S.\,S.~Kamalov, and L.~Tiator, Nucl.\ Phys.\
\textbf{A645}, 145 (1999) [nucl-th/9807001].

\bibitem{PDG}
Particle Data Group, Phys.\ Lett.\ B\textbf{592}, 1 (2004).

\bibitem{Chiang}
W.~T. Chiang, S.~N. Yang, L. Tiator, M. Vanderhaegen,
  and D. Drechsel, Phys.\ Rev.\ C~\textbf{68}, 045202 (2003).

\bibitem{others}
 B. Borasoy, Eur.\ Phys.\ J. \textbf{A9}, 95 (2000); B. Borasoy, E. Marco, and S.
Wetzel, Phys.\ Rev.\ C~\textbf{66}, 055208 (2002); A. Sibirtsev, Ch.\ Elster,
S. Krewald, and   J. Speth, nucl-th/0303044.

\bibitem{RSprop}
P.\,A.~Moldauer and K.\,M.~Case, Phys.\ Rev.\ \textbf{102}, 279 (1956); C.
Fronsdal, Nuovo Cimento Suppl.\ \textbf{9}, 416 (1958); A. Aurelia and
H.~Umezawa, Phys.\ Rev.\ \textbf{182}, 1682 (1969); L.\,M.~Nath, B. Etemadi,
and J.\,D.~Kimel, Phys.\ Rev.\ D\,\textbf{3}, 2153 (1971); M.~Benmerrouche,
R.\,M.~Davidson, and N.\,C.~Mukhopadhyay, Phys.\ Rev.\ C\,\textbf{39}, 2339
(1989).


\end{thebibliography}
\end{document}